\documentclass[%
 reprint,
 amsmath,amssymb,
 aps,
]{revtex4-2}
\usepackage{graphicx}
\usepackage{dcolumn}
\usepackage{bm}
\usepackage{bbm}
\usepackage[]{hyperref}

\hypersetup{colorlinks=true,linkcolor=blue,citecolor=blue,urlcolor=blue,pdfpagemode=UseNone}

\DeclareMathOperator{\diag}{diag}

\begin{document}

\title{Knight shift measurements probing Fermi surface changes under pressure in CeRhIn$_5$}

\author{Y.-H. Nian}
\author{C. Chaffey}
\author{P. Sherpa}
\author{L. Santillan}
\author{K. Nagashima}
\author{Peter Klavins}
\author{V. Taufour}
\author{N. J. Curro}
\affiliation{Department of Physics and Astronomy, University of California Davis, Davis, CA }

\date{\today}

\begin{abstract}
We report nuclear magnetic resonance (NMR) Knight shift measurements of the In(1) and In(2) sites in CeRhIn$_5$ as a function of pressure.  In contrast to the $c$ axis, the in-plane components of the In(1) Knight shift tensor exhibit little to no pressure dependence.  These results indicate that the dipolar component of the tensor is strongly suppressed at the In(1) site, while it remains constant with pressure at the In(2) site.  We analyze the hyperfine coupling in terms of a tight binding model for the electronic structure, and determine that the pressure dependence of the In(1) shift cannot be explained in terms of changes to the crystal field parameters, but rather can be understood in terms of an increase in the 4f electron content at the Fermi surface.  Our results indicate that the hyperfine coupling reflects changes in the electronic structure near a Kondo breakdown quantum critical point. 
\end{abstract}

\maketitle

\section{Introduction}

The basic physics of heavy electron materials has garnered the attention of the condensed matter community for decades, and continues to challenge our understanding of the limits of the Landau-Fermi liquid theory of metals \cite{doniach,zachreview,StewartHFreview}.  These materials contain a lattice of localized f-orbitals that hybridize with itinerant conduction electrons, and Coulomb repulsion at the f sites give rise to a range of strongly correlated electron behavior. Depending on the degree of hybridization, materials exhibit  localized f-electron magnetism mediated by Ruderman- Kittel-Kasuya-Yosida (RKKY) couplings, or Fermi liquid behavior with large effective mass carriers with Kondo screening of the f moments. Between these two extremes experiments have revealed evidence for a quantum phase transition, with a broad swath of non-Fermi liquid behavior extending to high temperatures \cite{Schroder2000, LohneysenQPTreview,Gegenwart2008, YRSdynamicalQCscaling}, and sometimes the presence of a dome of unconventional superconductivity \cite{ParkCeRhIn5PhaseDiagram,tusonNature2008}.  Importantly, Hall measurements and quantum oscillations studies indicate  that the Fermi surface appears to change dramatically at the quantum critical point (QCP) \cite{Loehneysen1996,Paschen2004, ShishidoRh115dHvA,Jiao2015,Chen2018b}, and theoretical studies suggest that the Fermi surface changes abruptly as the Kondo screening of the moments breaks down \cite{SiLocalQCP, ColemanHFdeath, Senthil2004,SiKondoDestruction2014JPSJ,Paschen2020,Hu2024,Gleis2024}.  On the other hand, Fermi surface measurements via angle-resolved photoemission (ARPES) studies have challenged this interpretation \cite{Kummer2015}.  Whether the superconductivity and and non-Fermi liquid behavior arise due to low energy fluctuations of the local moments at an antiferromagnetic QCP, or due to a Kondo breakdown QCP, remains a key open question \cite{Friedemann2009,Maksimovic2022}. 

CeRhIn$_5$ is an important archetype for investigating heavy fermion behavior across this quantum phase transition \cite{HeggerRh115discovery,Thompson2001,CeRhIn5ParkNJP2009}.  This material exhibits antiferromagnetic ordering below 3.8 K of partially-screened Ce f moments at ambient pressure \cite{baoCeRhIn5INS,Curro2000a}, and becomes superconducting at pressures above 1 GPa, reaching a maximum $T_c$ of 2.3 K near 2 GPa  \cite{CeRhIn5ParkNJP2009}.  Pressure is expected to change the hybridization between the f and conduction electrons, serving as a tuning knob for the Kondo interaction.  de Haas van Alphen measurements under pressure revealed an increase in the Fermi surface volume and a divergence of the effective mass at 2.35 GPa \cite{ShishidoRh115dHvA}.   Moreover, non-Fermi liquid behavior persists in the normal state  in a fan-like structure up to 10-15 K from a putative QCP at this pressure, which is hidden by the superconducting state \cite{tusonNature2008}.  A recent ARPES study at ambient pressure revealed a small Fermi surface, in contrast with those of the isostructural materials CeCoIn$_5$ and CeIrIn$_5$, where larger Fermi surfaces are believed to reflect the itinerant nature of the Ce f electrons \cite{Chen2018b}.  Nuclear magnetic resonance (NMR) experiments have probed the evolution of the spin fluctuations under pressure and uncovered power-law behavior of the spin lattice relaxation rate in the superconducting state \cite{Mito2001,kitaokaCeRhIn5pressureGapless}.  Recent Knight shift measurements suggested a change in the hyperfine couplings and the electric field gradient (EFG) under pressure \cite{Lin2015}.  The changes to the hyperfine couplings are similar to the differences in the couplings observed in CeCoIn$_5$ and CeIrIn$_5$, and have been attributed to modifications of the orbital anisotropy of  $J=5/2$ Ce crystal field  (CEF) ground state \cite{Willers_2015,Menegasso2021}.  However, the previous Knight shift measurements were conducted only for the field along the crystalline $c$ axis of the tetragonal unit cell, so the full tensor nature of the hyperfine coupling and its dependence on pressure has remained unknown.

\begin{figure*}
\centering
\includegraphics[width=\linewidth]{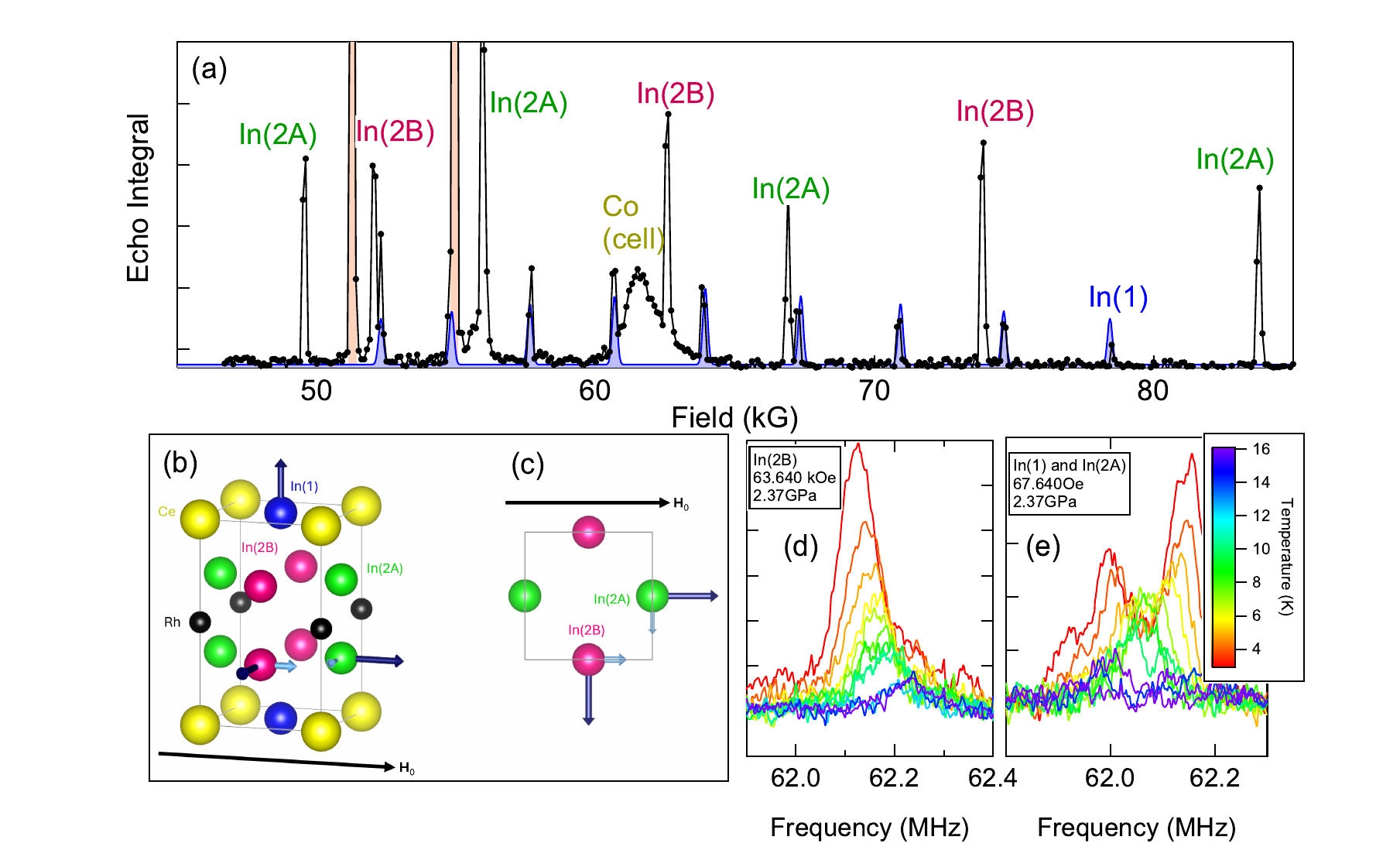}
\caption{\label{Fig:fieldsweep} (a) Field sweep at 62.15 MHz at 5 K and 0 GPa.  Blue line: fit to In(1) with $\omega_Q = 6.78$ MHz, $K = 2.7$\%, and $\theta = 81.3^{\circ}$. For In(2) $\omega_{zz} = 16.665$ MHz, $\eta = 0.445$. (b) CeRhIn$_5$ unit cell, with the EFG vectors, $\mathbf{q}$, for the In sites shown in dark blue, and the applied field, $\mathbf{H}_0$ shown in black.  The light blue arrows indicate the secondary EFG vectors, $\mathbf{q}_2$, as discussed in the text.  (c) The In(2) plane, indicating the difference between the In(2A) and In(2B) sites.  Frequency-swept spectra at constant field for the In(2B) (d), and In(1) and In(2A) (e) for several different temperatures at 2.37 GPa. 
}
\end{figure*}

Here we present $^{115}$In Knight shift measurements in CeRhIn$_5$ for field in the $ab$ plane.   We find that the Knight shift for In(1) site (in the Ce plane)  exhibits only a weak pressure dependence, in contrast with the behavior for the $c$ axis. The In(2) site exhibits little to no pressure dependence for any field direction.  These results indicate that both the dipolar and isotropic transferred hyperfine couplings between the f electron and the In(1) nuclear spins are both suppressed dramatically with pressure, whereas the coupling to the In(2) spins remain unaffected.  We analyze this behavior within a tight-binding model that includes the CEF terms, and find that neither changes to the crystal field nor changes to the hybridization can fully capture the observed trends.  Rather, the observations indicate that the fraction, $f$, of the 4f states at the Fermi surface increases as a function of pressure, consistent with a change of the Fermi surface across a Kondo breakdown QCP. More broadly, our results indicate that NMR measurements of the transferred hyperfine coupling are a sensitive measure of changes to the f electron localization in heavy fermion compounds. 

\begin{figure}
\centering
\includegraphics[width=\linewidth]{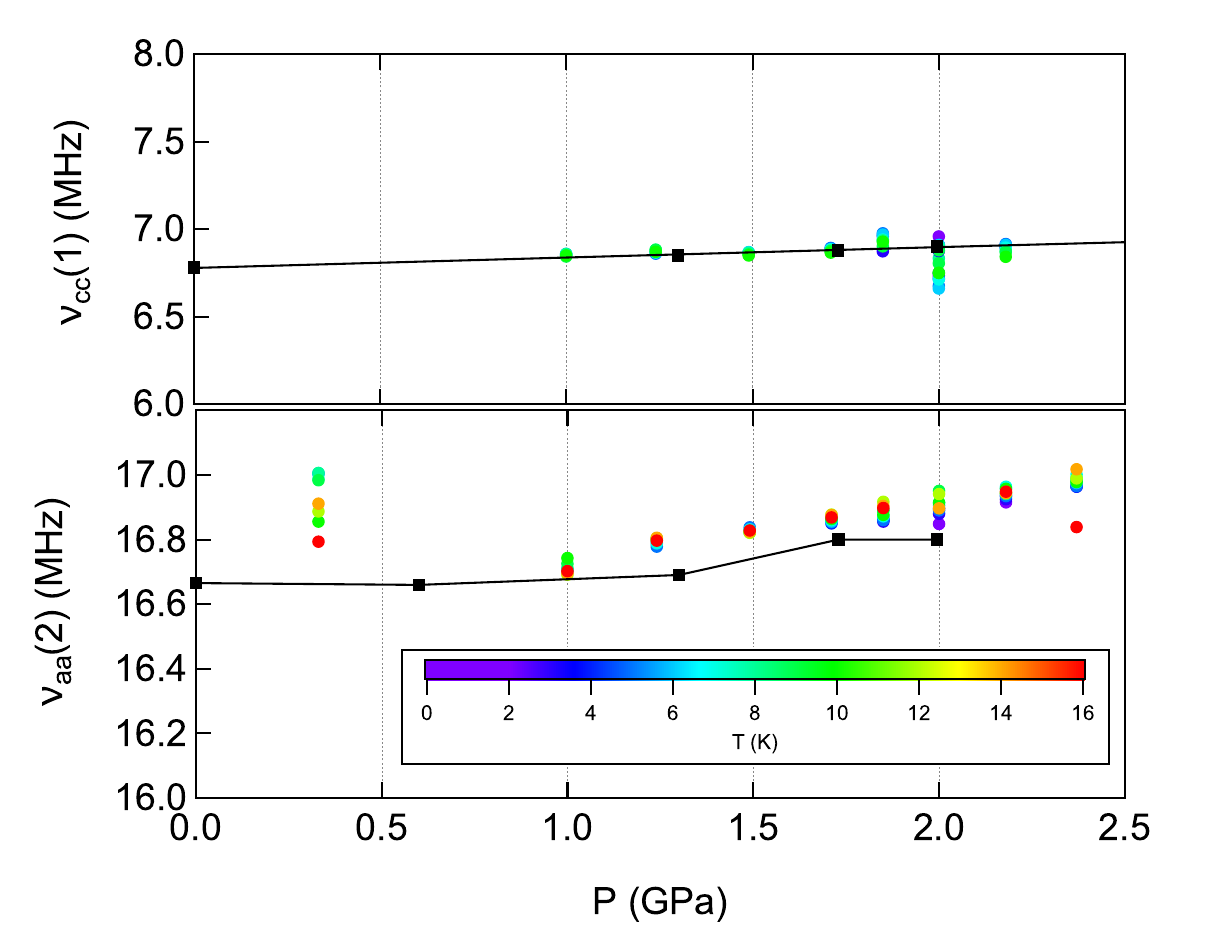}
\caption{\label{Fig:EFG} EFGs of the In(1) and In(2) sites versus pressure.  The colored circles are the measured data, and the solid black squares are reproduced from Ref. \cite{Lin2015}.  }
\end{figure}

\section{Experimental Results}

Single crystals were grown from In flux methods as described in \cite{HeggerRh115discovery}, and characterized by powder x-ray diffraction and specific heat measurements.  A crystal of approximate dimensions $2.0 \times 1.0\times 0.5$ mm$^{3}$ was polished and located in a coil inside of a  piston-cylinder pressure cell (easyLab P-cell 30), with Daphne oil as a pressure medium, such that the field is nominally aligned perpendicular to the $c$-axis.  Pressure was measured by ruby fluorescence \cite{Lin2015} at ambient temperature.  The NMR spectrum was measured by acquiring Hahn echos as a function of applied field at constant frequency, or as a function of frequency at constant field.  A typical spectrum is shown in Fig. \ref{Fig:fieldsweep}, which reveals $\sim 18$ independent resonances.

In order to discern which resonance is which, it is vital to understand the orientation of the applied field with respect to the crystal.  For the magnetic field $\mathbf{H}_0$ perpendicular to the $c$-axis, there are three distinct indium resonances.  Their frequencies are given by nuclear spin Hamiltonian:
\begin{equation}
    \mathcal{H}_{n} = \gamma\hbar \mathbf{I}\cdot(1 + \mathbb{K})\cdot\mathbf{H}_0 + \mathcal{H}_Q
\end{equation}
where $\gamma = 9.3925$ MHz/T is the gyromagnetic ratio, $\mathbb{K}$ is the Knight shift tensor, and the quadrupolar Hamiltonian is given by:
\begin{equation}
    \mathcal{H}_Q =  \sum_{\alpha\beta} \nu_{\alpha\beta}\left(\frac{3}{2}(I_{\alpha}I_{\beta} + I_{\beta}I_{\alpha}) -\delta_{\alpha\beta} I^2 \right)
\end{equation}
where $\nu_{\alpha\beta} = {eQV_{\alpha\beta}}/{6I (2I -1)}$ is diagonal in the crystal basis.  The EFG tensors differ for the two In sites, shown in Fig. \ref{Fig:fieldsweep}(b), and the resonances depend strongly on the field orientation.  It is thus possible to analyze the spectrum to discern which site is which, as detailed in Appendix \ref{app:sites}.  

\begin{figure}
\centering
\includegraphics[width=\linewidth]{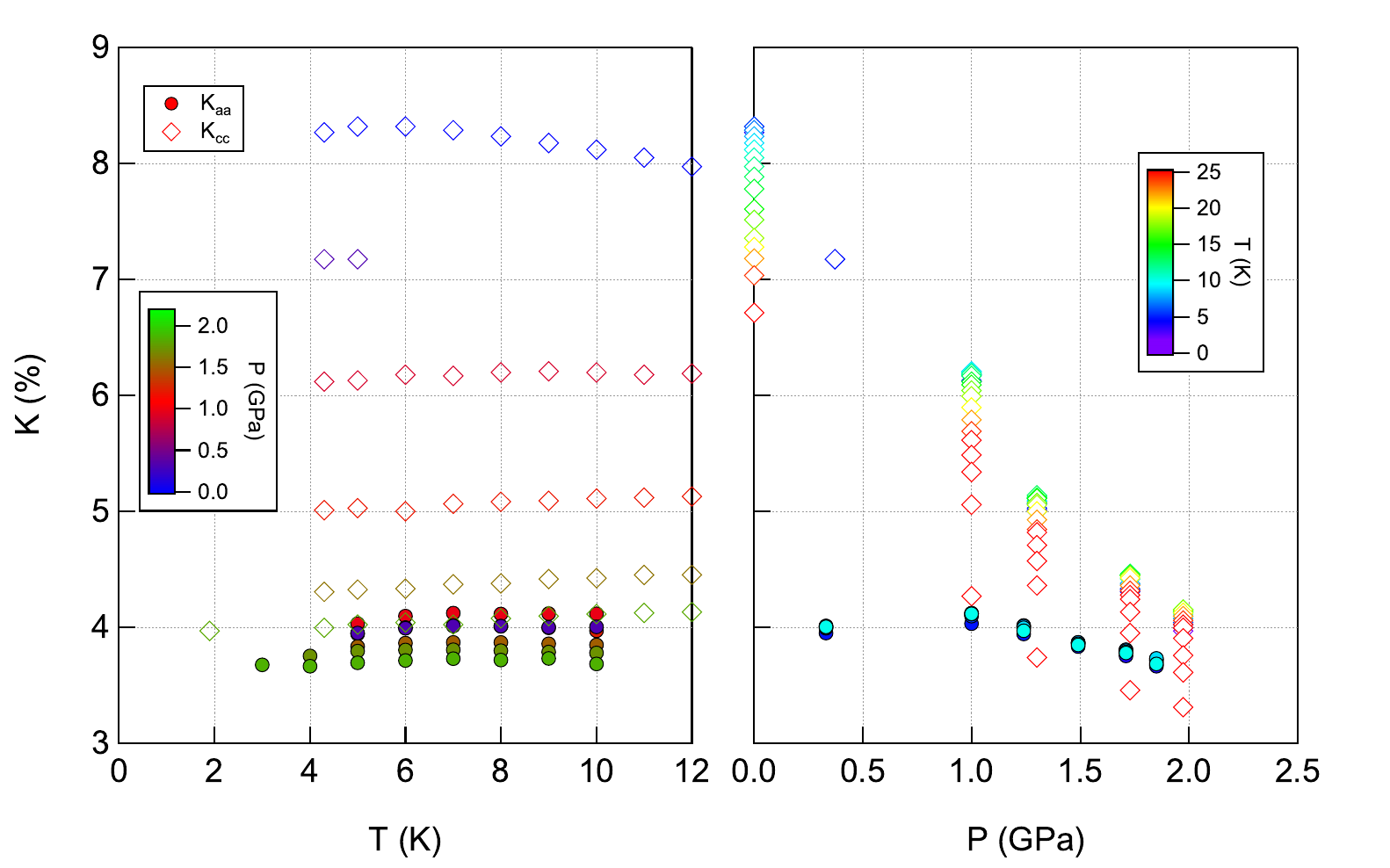}
\caption{\label{Fig:In1shift} The Knight shifts $K_{aa}$ and $K_{cc}$ of the In(1) site versus temperature and pressure.  The open diamonds ($K_{cc}$) are reproduced from Ref. \cite{Lin2015}.   }
\end{figure}

With the orientation fully determined at 0 GPa and 5 K, it is possible to identify each of the sites, as illustrated in Fig. \ref{Fig:fieldsweep}(a).  Several of these resonances were then measured by varying the frequency at fixed field for different temperatures at each pressure, as illustrated in Fig. \ref{Fig:fieldsweep}(d,e).  For the In(1) site there are two temperature and pressure-dependent quantities,  $K_{aa}(1)$ and $\nu_{cc}(1)$, which we extract from two such resonances by exploiting the relationship $f_n = \gamma H_0(1+K_{aa}) + n \nu_{cc}(3\cos^2\theta -1)/2$, where $\theta = 81.3^{\circ}$.   The pressure dependence of $\nu_{cc}$ matches that measured previously \cite{Lin2015}, as shown in Fig. \ref{Fig:EFG}, indicating that the orientation of the crystal did not change under pressure.  The Knight shift $K_{aa}(1)$ of the In(1) is shown in Fig. \ref{Fig:In1shift}, along with data for $K_{cc}(1)$ that is reproduced from Ref. \cite{Lin2015}.  Is is clear that the former exhibits only a modest variation with pressure, whereas the latter decreases by a factor of two over the same range of pressure. Both quantities exhibit a relatively flat temperature dependence. 


\begin{figure}
\centering
\includegraphics[width=\linewidth]{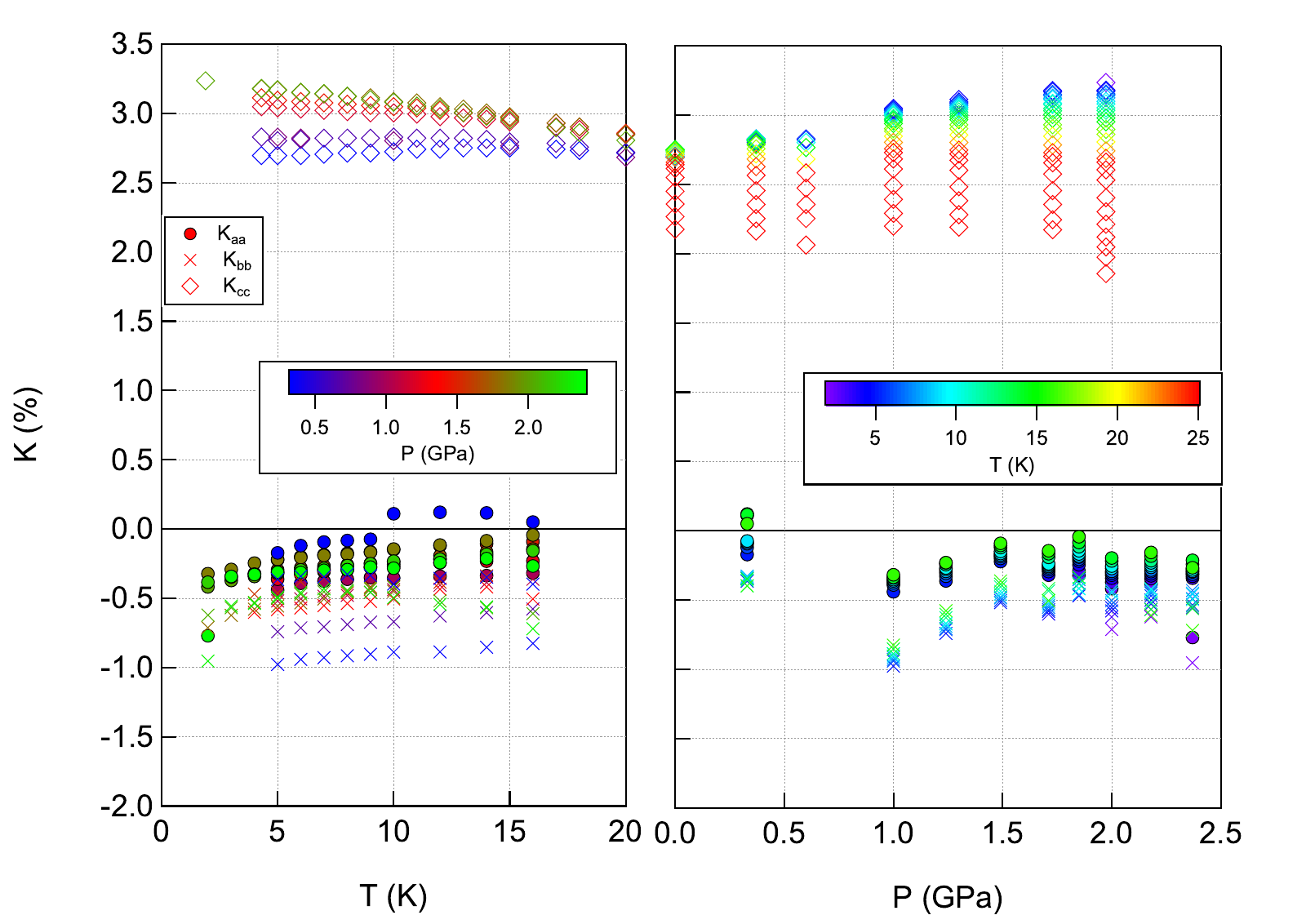}
\caption{\label{Fig:In2shift} The Knight shifts $K_{aa}$, $K_{bb}$ and $K_{cc}$ of the In(2) site versus temperature and pressure.  The open diamonds ($K_{cc}$) are reproduced from Ref. \cite{Lin2015}.   }
\end{figure}

For the In(2) site, there are four independent quantities that can be measured: $K_{aa}(2)$, $K_{bb}(2)$, $\nu_{aa}(2)$ and $\eta(2)$, however previous studies indicated that $\eta(2)$ is pressure independent.  We therefore measured one In(2A) and two In(2B) spectra to extract these quantities, assuming that $\eta(2)$ is constant.  Because the relationship between frequency and field is complicated for the In(2), we fit the calculated frequencies of these three transitions  to a linear expression, $f_i = a_i + b_i H (1+ K_{aa,bb}) + c_i\nu_{aa}$, which can then be inverted to extract the shifts and EFGs as a function of the measured  frequencies, $f_i$ and fields, $H$ (see Appendix \ref{app:sites} for details).  The EFG data is shown in Fig. \ref{Fig:EFG}.  The behavior is qualitatively similar to that measured previously, but is about 0.6\% higher.  Previous studies have suggested the presence of a discontinuity in the In(2) EFG near 1.5 GPa that has been intepreted as evidence for a QCP associated with a change in the Fermi surface \cite{Lin2015,Kawasaki2020}, however our data show only a gradual increase across this pressure.  Our assumption that $\eta$ remains constant may be responsible for this discrepancy. The Knight shifts, $K_{aa}$ and $K_{bb}$ are shown as a function of temperature and pressure in Fig. \ref{Fig:In2shift}.  The in-plane anisotropy of the shifts is approximately a factor of 3, and are about a factor of 5 smaller in magnitude than the shift in the $c$ direction.  In contrast to the In(1) site, there is a weak increase in $K_{cc}$ with pressure, with little to no change in $K_{aa}$ or $K_{bb}$ with pressure.  

\begin{figure}
\centering
\includegraphics[width=\linewidth]{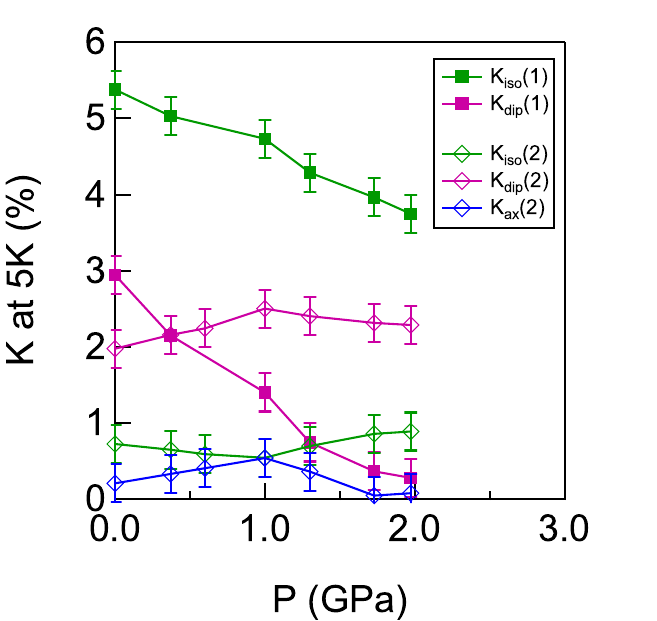}
\caption{\label{Fig:KvP} Knight shift components of the In(1) and In(2) sites at 5 K  versus pressure. }
\end{figure}

The Knight shift tensors can be decomposed into linear combinations of  isotropic, dipolar, and axial tensors:  $\mathbb{K} = K_{iso}\mathbf{1} + K_{dip} \mathbb{D} + K_{ax}\mathbb{X}$, where $\mathbf{1}$ is a the unit matrix, $\mathbb{D} = \diag(-1/2, -1/2, 1)$ and $\mathbb{X} = \diag(1/2, -1/2, 0)$.  Fig. \ref{Fig:KvP} shows how the $K_{iso}$, $K_{dip}$ and $K_{ax}$ vary as a function of pressure at 5K for the two In sites.  $K_{ax}$ vanishes for the high symmetry In(1) site, but $K_{iso}$ and $K_{dip}$ are suppressed by pressure such that the shift becomes almost completely isotropic by 2 GPa.  The In(2) site is dominated by the dipolar term, but shows little to no pressure dependence. The pressure dependence at other temperatures (not shown) is similar.


\section{Discussion}

The temperature and pressure dependence of the Knight shift are determined by the hyperfine couplings and the spin susceptibilities of the different spin components.  In general $\mathbb{K} = \mathbb{K}_0+\sum_{i,j}\mathbb{A}_i\cdot\chi_{ij}$, where $\mathbb{K}_0$ is a constant, $\mathbb{A}_i$ is a hyperfine coupling tensor to spin $\mathbf{S}_i$ and $\chi_{ij} = \langle \mathbf{S}_i\mathbf{S}_j\rangle$.  If there is more than one spin degree of freedom, such as in the heavy fermions, then temperature dependence of the Knight shift can be complicated, depending on the relative scale of the different hyperfine coupling constants.  A previous study indicated that $\chi_c$ remained relatively unchanged under pressure in CeRhIn$_5$, suggesting that the pressure dependence is explained by changes to the hyperfine coupling \cite{Lin2015}. The hyperfine constants depend on the local bonding and electronic structure, however little is known about how these couplings behave in heavy fermion materials.  

\subsection{Transferred Hyperfine Coupling}

The hyperfine interaction between a nucleus at the origin and an electron at located at position $\mathbf{r}$ is given by:
\begin{equation}
\label{eqn:hyperfine}
    \mathcal{H}_{h} = g\mu_B\gamma\hbar\mathbf{I}\cdot\left(\frac{\mathbf{L}}{r^3} +3\frac{\mathbf{r}(\mathbf{S}\cdot\mathbf{r})}{r^5}-\frac{\mathbf{S}}{r^3}+\frac{8}{3}\pi \mathbf{S} \delta(\mathbf{r})\right),
\end{equation}
where $\gamma$ is the nuclear gyromagnetic ratio, $\mu_B$ is the Bohr magneton, $g$ is the electron $g$-factor, and $\mathbf{L}$ is the angular momentum relative to the nucleus \cite{abragambook}.  This interaction must be summed over all the unpaired electron spins in a solid. The last term, known as the Fermi contact interaction, is typically the largest. It arises from s orbitals and gives rise to an isotropic coupling $A_{iso}\mathbf{I}\cdot\mathbf{S}$ where  $A_{iso} \sim 10^2-10^3$ kOe/$\mu_B$.   If the electron is within a non-s orbital at the nuclear site, the hyperfine interaction has a dipolar form: $\mathcal{H}_{h}=\mathbf{I}\cdot \mathbb{A}_{dip}\cdot\mathbf{S}$, where $\mathbb{A}_{dip}$ is a traceless tensor. 
For an electron located on a different atom than the nucleus in question, $\mathbb{A}_{dip}$ reduces to the classical direct dipole interaction.   In solids the magnitude of the direct dipolar coupling is  $\sim 0.1-0.5$ kOe/$\mu_B$, much smaller compared with on-site couplings, and can often be neglected. In materials where the orbitals hybridize, however, the hyperfine coupling between a nucleus at the origin and an electron on a distant orbital can reach values comparable with $A_{iso}$, and is known as \emph{transferred} hyperfine coupling \cite{Owen1966}.    
 
In heavy fermions, the hyperfine interaction at a ligand site such as the In in CeRhIn$_5$ is often modeled with two sets of electronic spins: an on-site coupling, $\mathbb{A}$,  to the itinerant conduction electrons, $\mathbf{S}_c$, and a transferred coupling, $\mathbb{B}$, to spins on the 4f orbitals, $\mathbf{S}_f$ \cite{Curro2004,ShirerPNAS2012}:
\begin{equation}
\label{eqn:twobands}
    \mathcal{H}_{h} =  g\mu_B\gamma\hbar\mathbf{I}\cdot\left(\mathbb{A}\cdot\mathbf{S}_c +  \mathbb{B}\cdot\mathbf{S}_f\right).
\end{equation}
The transferred couplings can be determined by comparing the Knight shift and bulk susceptibility as a function of temperature and field direction, and have been well-documented for the stoichiometric CeMIn$_5$ materials, as summarized in Table \ref{tab:table1}. Surprisingly, the transferred coupling $B_{cc}(1)$ for the In(1) site decreases by a factor of three between M=Rh to M=Co, whereas $B_{cc}(2)$ for the In(2) site increases by the same factor. Evidence for a similar evolution of $B_{cc}(1)$ has been observed in CeRhIn$_5$ under hydrostatic pressure  \cite{Lin2015}. Such a large variability in transferred hyperfine couplings constants has not been observed in other strongly correlated superconductors, such as the cuprates, iron pnictides or chalcogenides \cite{mila89,T1formfactorsArsenides},  and there has been little to no work to understand the transferred hyperfine coupling in the heavy fermions quantitatively. There are two viewpoints that have been discussed in the literature concerning the  variability among these constants. 

\begin{table}[b]
\caption{\label{tab:table1}%
Transferred hyperfine coupling constants (in kOe/$\mu_B$) for the CeMIn$_5$ materials at ambient pressure. Reproduced from \cite{Curro2004, ShirerPNAS2012,CeIrIn5HighFieldNMR}.
}
\begin{ruledtabular}
\begin{tabular}{lddd}
&
\multicolumn{1}{c}{\textrm{CeRhIn$_5$}}&
\multicolumn{1}{c}{\textrm{CeIrIn$_5$}}&
\multicolumn{1}{c}{\textrm{CeCoIn$_5$}}\\
\colrule
$A_{aa}(1)$ & 19.6 & 11.3 & 12.1 \\
$A_{cc}(1)$ & 21.45 & 13.8  & 8.9\\
$A_{aa}(2)$ & - & - & -0.4\\
$A_{bb}(2)$ & - & - & 10.3\\
$A_{cc}(2)$ & - & 20.8 & 28.1\\
\end{tabular}
\end{ruledtabular}
\end{table}
   
\emph{Crystal field model}:   The variability between compounds has previously  been attributed to changes in the shape of the Ce 4f ground state orbital \cite{Shockley2015,Menegasso2021}.  X-ray absorption spectroscopy measurements indicate that the orbital anisotropy evolves across the CeMIn$_5$ series \cite{Willers_2015}.  The anisotropy can be characterized by $\alpha^2$, the fraction of the $|J_z=\pm 5/2\rangle$ state in the ground state wavefunction of an isolated Ce ion. When $\alpha^2$ is large, the ground state orbital is more spatially extended along the Ce-In(1) direction, suggesting a larger degree of orbital overlap and hence a larger hyperfine coupling. The orbital anisotropy is also a strong function of the CEF parameters, and may correlate with the $c/a$ ratio  \cite{Pagliuso2002,Pagliuso2002c}.

\emph{Hybridization model}: An alternative explanation is that differences in hybridization between the Ce and In orbitals are responsible for the variation in transferred hyperfine couplings. 
A finite hybridization, $V$,  between f and p orbitals can leave a fraction, $f$,  of unpaired spin in the f orbital.  In this case the hyperfine coupling to the ligand nucleus is $(1-f) \mathbb{A}_0$, where $\mathbb{A}_0={4}g\mu_B \langle r^{-3}\rangle/{5}$ is the coupling corresponding to the ligand p orbital being singly occupied  \cite{Owen1966, abragambook}.  
In a metal, only the states at the Fermi level contribute to the hyperfine interaction: for the states below both spin states are occupied and therefore the hyperfine field cancels out, and the states above are unoccupied and do not contribute. It follows that if the states at $E_F$ have more f than p character, then $1-f$ will be reduced and the hyperfine coupling should be smaller.   On the other hand, for a simple model of hybridized orbitals, $1-f = (V/\Delta)^2$, where $\Delta$ is the difference in energy of the Ce f and p atomic states  \cite{Owen1966}.   This result would suggest that an increase in hybridization should increase the hyperfine coupling.

\subsection{Dependence on electronic Hamiltonian}

\begin{figure}
\centering
\includegraphics[width=\linewidth]{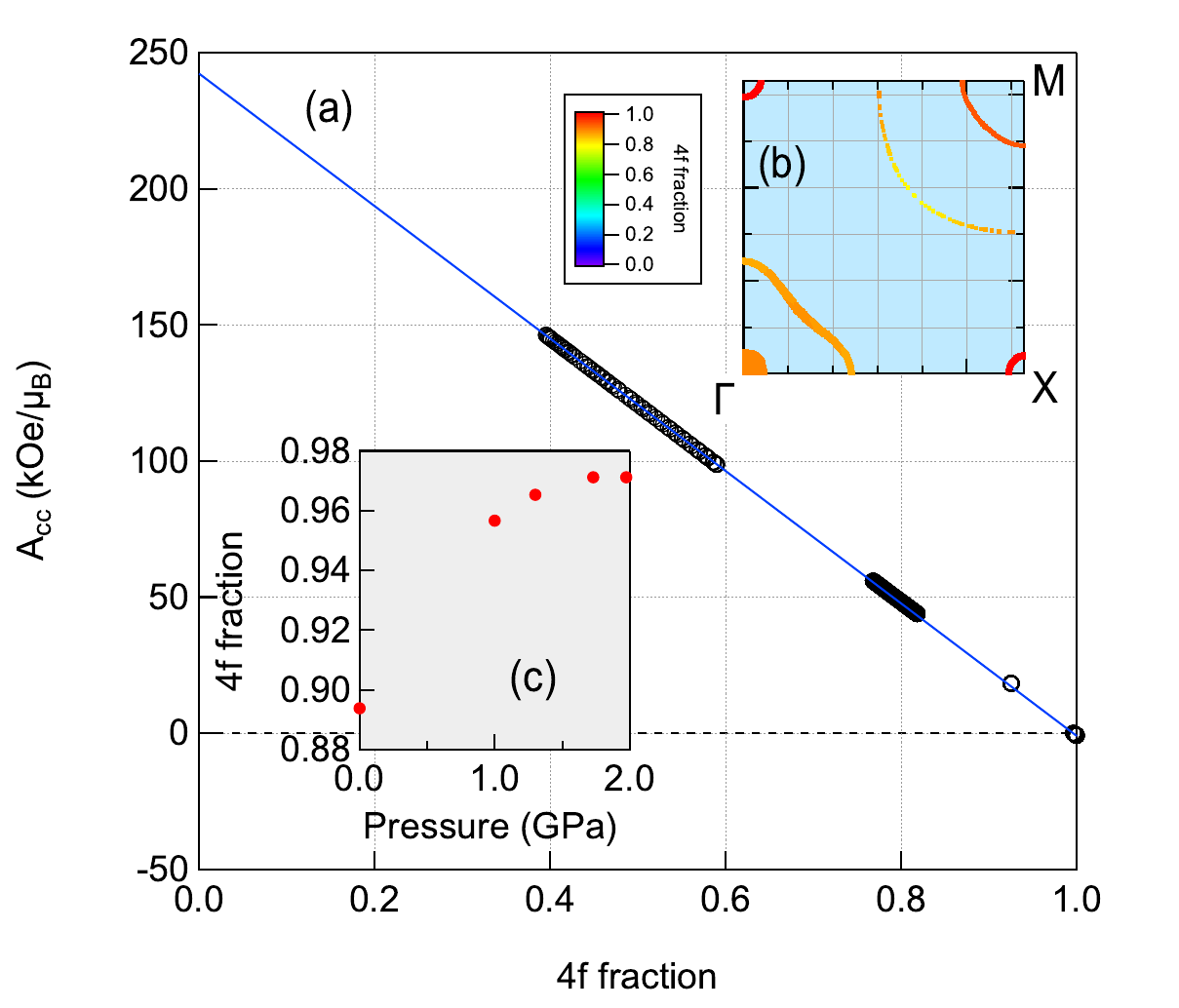}
\caption{\label{fig:hypVfrac} (a) Hyperfine coupling $A_{cc}$ versus fraction of 4f states on the Fermi surface. The solid line is a linear fit as described in the text.  (b) The Fermi surface computed using the tight binding model, and the color bar indicates the fraction of 4f states. (c) The 4f fraction versus pressure derived from the hyperfine coupling.  }
\end{figure}

To better understand the effects of the CEF and hybridization terms on the hyperfine interaction, we treat  $\mathcal{H}_{h}$ in Eq. \ref{eqn:hyperfine} as a first-order perturbation  and compute the matrix elements $\langle \psi|\mathcal{H}_{h}|\psi'\rangle$.  For simplicity, we adopt a minimal tight-binding model for the electronic states, $|\psi\rangle$, in the CeMIn$_5$ system developed by Maehira and coworkers  \cite{Maehira2003}:
\begin{equation}
\label{eqn:tightbinding}
    \mathcal{H}_e = \mathcal{H}_f + \mathcal{H}_p + \mathcal{H}_{fp} + \mathcal{H}_{CEF}.
\end{equation}
Here $\mathcal{H}_f$, $\mathcal{H}_p$ and $\mathcal{H}_{fp}$ represent Ce-Ce, In-In, and Ce-In hybridization, respectively, and are given in Appendix \ref{app:calc}.  These three terms are scaled by the hopping terms $V_{ff}$, $V_{pp}$ and $V_{pf}$, and $\Delta$ characterizes the difference in energy between the Ce and In atomic energies. $\mathcal{H}_{CEF}$ is the crystal field interaction, characterized by the Steven's operators with coefficients $B_{2}^0$, $B_{4}^0$ and $B_{4}^4$.    This model captures relevant features of the Fermi surface of CeCoIn$_5$ and CeIrIn$_5$ with only the Ce and In(1) atoms in a 2D lattice, and  importantly includes the CEF interaction. The tight-binding parameters necessary to describe CeRhIn$_5$, however, are not known. The model includes five orbitals: three from the Ce 4f and two from the In 5p orbitals, consisting of a Hilbert space of dimension 10, including spin.   Note that  this model does not include Coulomb interactions, which are important to important to capture the essential elements of the electron correlations and the dramatic changes to the Fermi surface at the QCP.  Our aim, however, is only to examine qualitatively how the hyperfine couplings at the In(1) site evolve as a function of the electronic parameters rather than to find quantitative agreement.  For example, this model does not contain $s$ orbitals and ignores any  core polarization effects \cite{abragambook}.   The hyperfine couplings, $A_{\alpha\beta}$, are determined by expressing $\langle \mathcal{H}_h \rangle = \gamma\hbar \mathbf{I}\cdot \mathbf{h}$, where $\mathbf{h}= \mathbb{A}\cdot\mathbf{S}$ is the hyperfine field of the electronic states averaged over the Fermi surface, and $\mathbf{S}$ is the spin of the electron (see Appendix \ref{app:calc} for details). $\mathbb{A}$ has dipolar symmetry, with the largest eigenvector along the $c$ axis, and is thus fully characterized by $A_{cc}$. Note that there are three Fermi surfaces in this model, each of which contains some fraction of the 4f orbitals (see inset of Fig. \ref{fig:hypVfrac}), thus it is not possible to decompose $\mathcal{H}_h$ in terms of on-site and localized spins, as in Eq. \ref{eqn:twobands}.  The three Fermi surfaces arise from the three pairs of $J=5/2$ states of the Ce that hybridize with one another and the In 5p states, and thus $A_{cc}$ indeed represents a transferred hyperfine coupling between the nucleus (at $\mathbf{r}=0$) and the Ce spin degrees of freedom.

We first consider the relationship between the hyperfine coupling and the nature of the states at the Fermi surface.  
Fig. \ref{fig:hypVfrac} displays $A_{cc}$ for different states at $E_F$ as as function of the fraction, $f$, of the 4f content for each of these electronic states.  The solid line is a linear interpolation between $A_{cc}^0$, the hyperfine coupling for a singly occupied 5p state, and $A_{cc}^{f}$, the direct dipolar coupling for an electron exclusively localized at the 4f site. 
The linear relationship indicates that for each electronic state, $A_{cc}(\mathbf{k}) = A_{cc}^0 + (A_{cc}^f - A_{cc}^0) f(\mathbf{k}) $.  This trend agrees with the hybridization model for transferred hyperfine coupling, and suggests that the coupling strength is a direct measure of the fraction $f$.  
The fact that $A_{cc}$ for the In(1) site  decreases with pressure in CeRhIn$_5$ indicates that the 4f content at the Fermi surface(s) that couple to the In(1) increases, independent of any particular electronic model.  

\begin{figure}
\centering
\includegraphics[width=\linewidth]{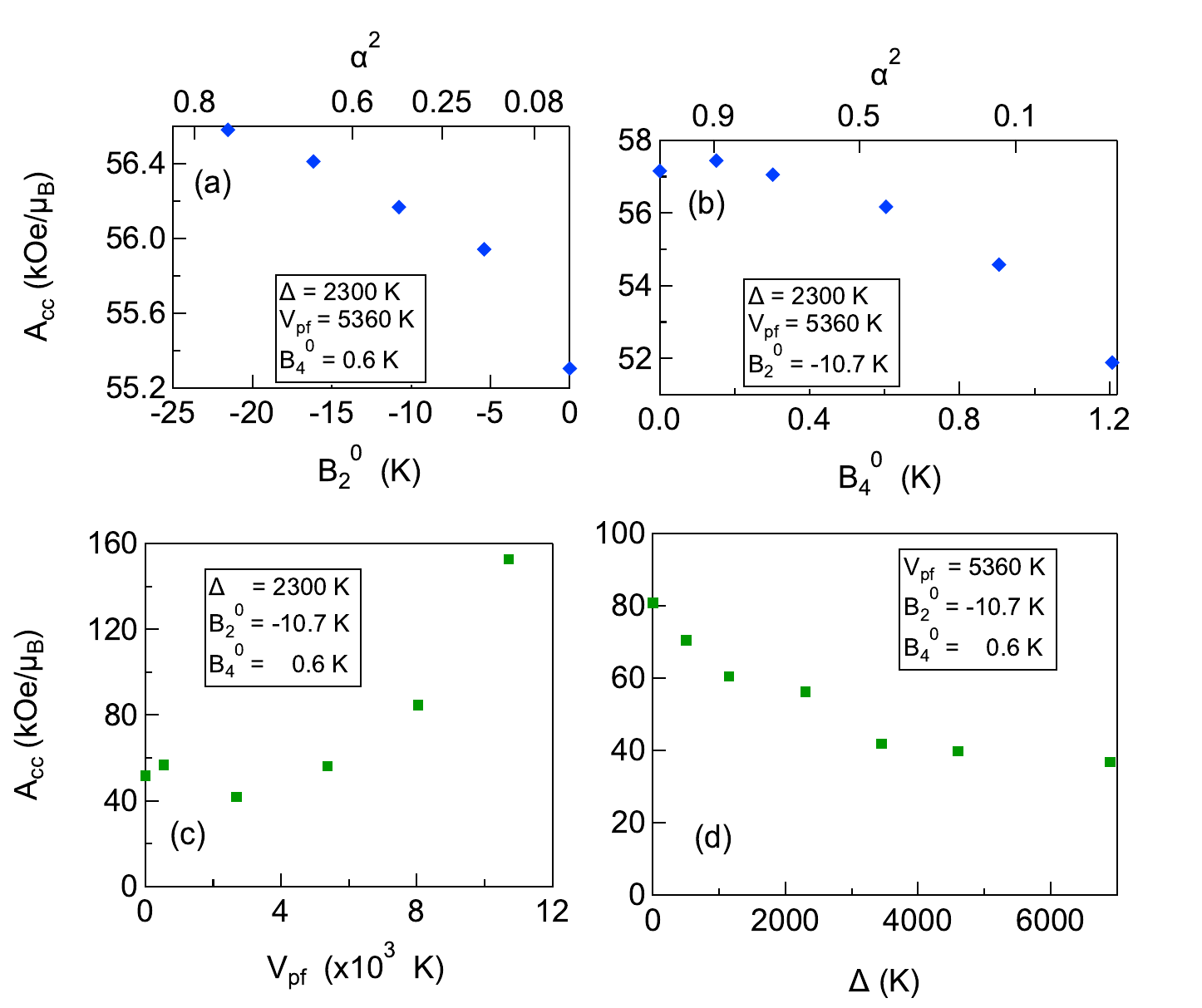}
\caption{\label{fig:HyperfineCEF} Hyperfine coupling $A_{cc}$ to the In(1) site as a function of (a) the crystal field parameters $B_2^0$, (b) the crystal field parameters $B_4^0$, (c) the In-Ce hybridization $V_{pf}$, and (b) the energy difference, $\Delta$, between the Ce and In atomic levels.  The orbital anisotropy coefficient, $\alpha^2$, is shown on the upper axes of panels (a) and (b).}
\end{figure}

We next examine the effect of changing the CEF parameters that affect the orbital anisotropy. $\mathcal{H}_{CEF}$ is much smaller than the hybridization terms in Eq. \ref{eqn:tightbinding}, so it is unclear how much a change in $\alpha^2$ would actually change $|\psi\rangle$ and hence $\mathcal{H}_h$. Figure \ref{fig:HyperfineCEF}(a) shows how $A_{cc}$, averaged over the Fermi surface, changes as a function of the CEF parameter $B_{2}^0$ within the tight-binding model calculation.  The anisotropy parameter is shown in the upper axis.  $B_2^0$  quantifies the tetragonal component of crystal field potential, and thus may correlate with the $c/a$ ratio  \cite{Pagliuso2002,Pagliuso2002c}.  However, as shown in Fig. \ref{fig:HyperfineCEF}(a), $A_{cc}$ varies by only 2\% as $B_2^0$ changes, even though the orbital anisotropy changes dramatically over this range.  This trend does not match the observed changes in the hyperfine coupling at the In(1) site,  suggesting that $B_2^0$ has little to no effect on the hyperfine coupling. 

Fig. \ref{fig:HyperfineCEF}(b) shows how $A_{cc}$ changes with $B_4^0$.  Changes in $B_4^0$ tend to narrow the bandwidth near $E_F$, and it has been suggested that this quantity can be tuned by the MIn$_2$ layer \cite{Maehira2003,Sundermann2016}.  In this case, $A_{cc}$ is reduced by 11\% as $B_4^0$ changes from 0 to 1.2 K, or as $\alpha^2$ is reduced from nearly unity to close to zero.  Although changing $B_4^0$ affects $A_{cc}$ more than changing $B_2^0$ does, the overall change is smaller than observed experimentally, and in the opposite direction.  In fact, CeRhIn$_5$ has the largest $B_4^0$ and also the largest $A_{cc}$, whereas CeCoIn$_5$ has the smallest $B_2^0$ and $A_{cc}$.  


We also examine the dependence on hybridization, $V_{pf}$, and the gap $\Delta$, shown in Fig. \ref{fig:HyperfineCEF}(c,d).  Increasing $V_{pf}$ increases $A_{cc}$  in an almost quadratic fashion, and increasing the energy difference, $\Delta$, decreases the coupling strength.  These trend agrees qualitatively with the argument that the transferred hyperfine coupling should vary as $(V_{pf}/\Delta)^2$ \cite{Owen1966}.  However, these changes just reflect changes to the average fraction $\langle f(\mathbf{k})\rangle_{E_F}$ over the Fermi surface. Increasing $V_{pf}$ in this model will decrease $f$ (and enhance the hyperfine coupling) by increasing the admixture of 5p states, whereas increasing $\Delta$ will have the opposite effect.   The fact that this simple single-electron model is unable to capture the observed trends in the hyperfine coupling signifies that more complicated physics is at play. Of course, this model does not account for any correlations or Coulomb interactions, which are responsible for the dramatic changes to the Fermi surface expected in the periodic Anderson model \cite{Gleis2024}. 

If we assume that the dominant contribution to the In(1) Knight shift pressure dependence arises from changes to the fraction, $f$, of the 4f electrons at the Fermi level, then it is possible to extract a rough estimate of $f$ versus pressure, as illustrated in Fig. \ref{fig:hypVfrac}(c).  Here we assume that the hyperfine coupling is solely due to the dipolar field from the In 5p electrons, as captured by the tight-binding model.  In reality, there will likely be other contributions arising from other channels, for example between the s orbitals of the In(1) and the Ce, as well as core-polarization contributions.  Nevertheless, there is a clear increasing trend of 4f contribution with pressure.   This observation agrees qualitatively with Quantum Monte Carlo (QMC) and Dynamical Mean Field Theory (DMFT) calculations in compressed Ce \cite{McMahan2003} and in the periodic Anderson model \cite{Gleis2024}, that indicate $f$ increases with pressure.  

In summary, we find that the changes to the dipolar component of the In(1) Knight shift, $K_{dip}(1)$ (Fig. \ref{Fig:KvP}), are likely driven by a pressure dependent hyperfine coupling that reflects changes to $f$, the 4f fraction of states at the Fermi level, which signal a change from small to large Fermi surface across a Kondo breakdown QCP.   Similar behavior is also seen in the isotropic component, $K_{iso}(1)$, which likely reflects a transferred hyperfine coupling to the In 6s orbitals, but is not captured within the tight-binding model.  The observed variation in the hyperfine coupling cannot be understood in terms of changes to the CEF parameters, $B_2^0$ and $B_4^0$.   The In(2) sites show little to no pressure dependence.  This behavior may signify that the In(2) couples to different Fermi surfaces that are already strongly hybridized at ambient pressure \cite{HauleCeIrIn5}.  

An important outcome of this work is that the hyperfine coupling can be easily understood in terms of the fractional content of the ligand site in the electronic states.  This approach eliminates the need for complicated calculations involving the expression in Eq. \ref{eqn:hyperfine}.  The hyperfine coupling is thus a sensitive probe that can reveal subtle changes in $f$ which may not be observable via more traditional techniques such as X-ray analysis \cite{Sundermann2016}.  Future experiments to measure the hyperfine coupling in other heavy fermion systems, such as YbRh$_2$Si$_2$ \cite{Paschen2016}, may help shed important light on how Fermi surfaces may change as the Kondo effect breaks down.

\begin{acknowledgments}
We acknowledge stimulating discussions with W. Pickett and R. Scalettar. Work at UC Davis was supported by the NSF under Grant No.  DMR-2210613. We acknowledge support from the Physics Liquid Helium Laboratory Fund.  
\end{acknowledgments}

\appendix

\section{Site Identification \label{app:sites}}

$^{115}$In has spin $I=9/2$, thus there are multiple resonances for each site, as illustrated in Fig. \ref{Fig:fieldsweep}(a).  The different In sites have distinct EFGs that enable identification of each of the resonances.  The EFG tensor is given by: $\nu_{\alpha\beta} = \diag(\nu_{aa}\nu_{bb}, \nu_{cc})$. The principal direction with the largest eigenvalue is known as the EFG vector, $\mathbf{q}$, and here we identify the axis with the second largest eigenvalue as the secondary EFG vector, $\mathbf{q}_2$.  For the In(1) site at ambient pressure,  $\nu_{aa} = \nu_{bb} = -\nu_{cc}/2 = -3.39$ MHz and $\mathbf{q} \parallel c$, as illustrated in Fig. \ref{Fig:fieldsweep}(b).  Because this site has axial symmetry, the asymmetry parameter $\eta = 0$ and there is no secondary EFG vector.   A magnetic field $\mathbf{H}_0 \perp c$ breaks the symmetry of the four In(2) sites. For the In(2A) at ambient pressure,  $\nu_{aa} =\nu_{zz} =16.665$ MHz, $\nu_{bb} = -(1+\eta)\nu_{zz}/2$ and $\nu_{cc}=-(1-\eta)\nu_{zz}/2$, where $\eta = 0.445$ \cite{Curro2000a}.  In this case, $\mathbf{q} \parallel a$ and $\mathbf{q}_2 \parallel b$, as illustrated in Fig. \ref{Fig:fieldsweep}(b,c).  For the In(2B) site, related by a $90^{\circ}$ rotation, $\nu_{aa}$ and $\nu_{bb}$ are switched (corresponding to $\eta \rightarrow -\eta$), as are $\mathbf{q}$ and $\mathbf{q}_2$.

The anisotropy of the EFG tensors means that the In resonance frequencies are strong functions of the field orientation relative to the crystalline axes. Moreover, because there are so many satellite transitions, it is not straightforward to identify which transition is which. The spectrum in Fig. \ref{Fig:fieldsweep}(a) reveals a series of nine peaks that are nearly equally spaced, corresponding to the In(1) site.  The filled blue region corresponds to a fit with $\mathbf{H}_0$ oriented $\theta = 81.3^{\circ}$ from the In(1) EFG vector along the $c$-axis.  

\begin{figure*}
\centering
\includegraphics[width=\linewidth]{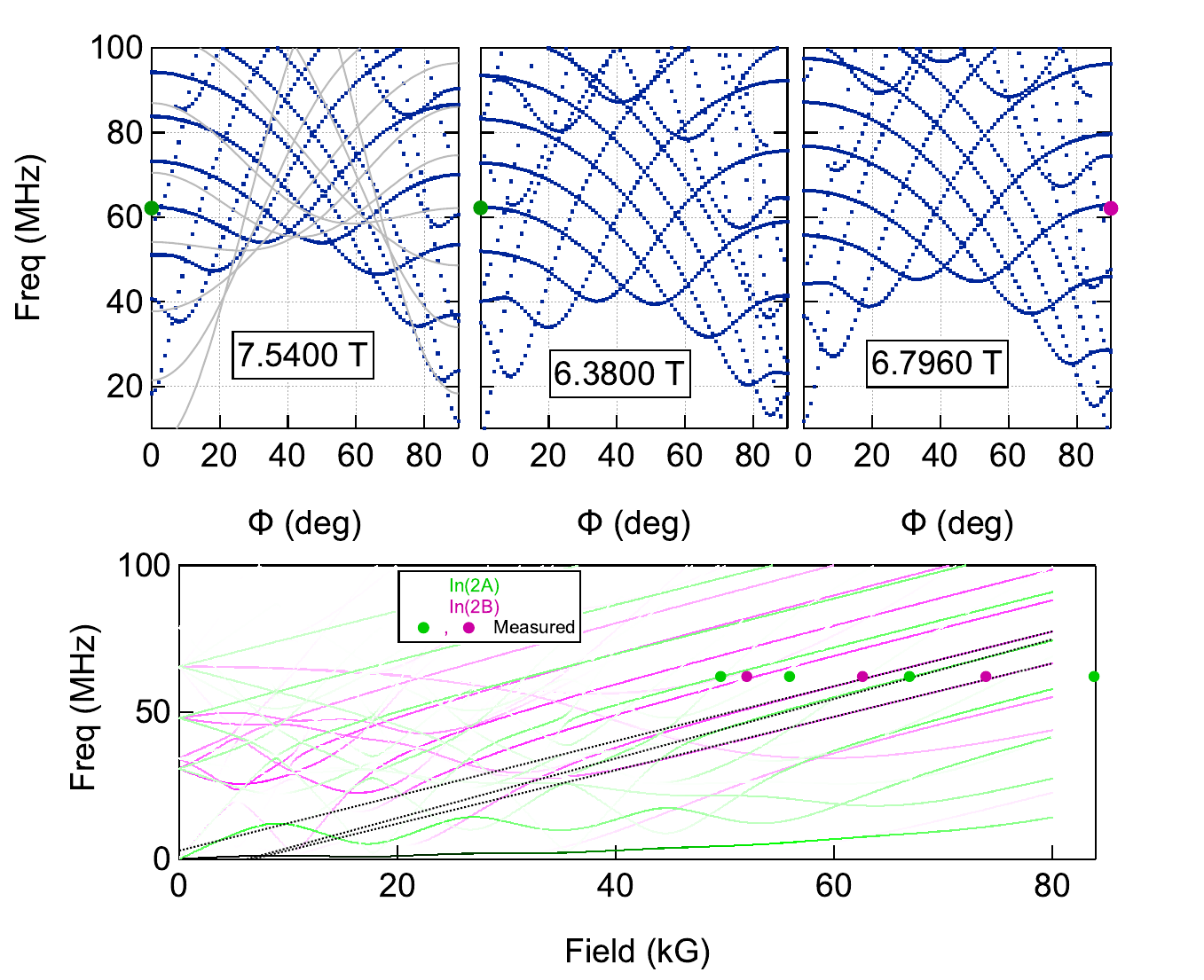}
\caption{\label{Fig:rotatephi} Upper row: Exact diagonalization calculations of frequency versus $\phi$ for three different fields at a fixed $\theta = 81.3^{\circ}$. The solid points are measured at 5K and 1 GPa.  The solid gray lines in the left panel are calculated using second order perturbation theory, but do not accurately represent the resonance frequencies for all angles. Lower panel: The frequency versus field relationship for the In(2A) and In(2B) sites.  The solid points correspond to the resonances shown in panel (a). The dotted lines are linear fits as described in the text.  }
\end{figure*}

Identifying the In(2) sites is more complicated.  In principle, $\mathbf{H}_0$ can lie anywhere on a cone around the $c$ axis with angle $\theta$, whereas the resonance frequency depends critically on the orientation of $\mathbf{H}_0$ relative to both $\mathbf{q}$ and $\mathbf{q}_2$.  Fig. \ref{Fig:rotatephi} shows how the In(2) resonances depend on the azimuthal angle $\phi$ between the projection of $\mathbf{H}_0$ on the $ab$ plane and the crystalline $a$ axis at fixed $\theta = 81.3^{\circ}$ for several different magnetic fields.  For $\phi = 0$, $\mathbf{H}_0$ is parallel  (or close to) $\mathbf{q}$ and we refer to these sites as In(2A).  For $\phi = 90^{\circ}$ $\mathbf{H}_0 \parallel \mathbf{q}_2$, and we refer to these as In(2B). The measured resonance frequencies at the three fields are indicated as solid points in Fig. \ref{Fig:rotatephi} and indicate that $\phi = 0$.    Figure \ref{Fig:fieldsweep}(d) shows how the resonance fields of the two In(2) sites varies as a function of the magnitude of $\mathbf{H}_0$ at this angle, as determined by exact diagonalization of $\mathcal{H}_{n}$.   For low to intermediate values of field and frequency, the resonances grow non-monotonically with field due to the complex interplay between the Zeeman and quadrupolar terms in $\mathcal{H}_{n}$.   The solid points indicate the various In(2A) and In(2B) resonances shown in  Fig. \ref{Fig:fieldsweep}(a).  The light gray lines in the left panel of Fig. \ref{Fig:rotatephi} show the result of second order perturbation theory in $\mathcal{H}_Q$. It is clear that such an approximation is insufficient for the In(2) site.

\section{Computation of Hyperfine Couplings  \label{app:calc}}

The hyperfine field can be expressed as:
\begin{equation}
    \mathbf{h} = g\mu_B\left(\frac{\mathbf{r}\times\mathbf{p}}{r^3} +3\frac{\mathbf{r}(\mathbf{S}\cdot\mathbf{r})}{r^5}-\frac{\mathbf{S}}{r^3}\right).
\end{equation}
The first term is usually written in terms of $\mathbf{L} =\mathbf{r}\times\mathbf{p}$, but we will consider the case where the origin is not centered at the atomic site.  The contact term is not included because no s-orbitals are involved.    We consider the matrix elements  within each spin-degenerate band: 
$\langle \mathbf{k}, n, s|{h}_{\alpha}|\mathbf{k}, n, s'\rangle$,
where $\alpha = a,b,c$ and $|\mathbf{k}, n, s\rangle$ is a Bloch state with wavevector $\mathbf{k}$, band index $n$, and spin $s$. These can be expanded as: 
\begin{equation}
    \langle \mathbf{k}, n, s|{h}_{\alpha}|\mathbf{k}, n, s'\rangle =  
    \zeta\mathbf{1} +A_{\alpha x}\sigma_x +A_{\alpha y}\sigma_y + A_{\alpha z}\sigma_z,
\label{eqn:hypmatrix}
\end{equation}
where   $A_{\alpha\beta}$ are hyperfine matrix elements, and $\sigma_{\alpha}$ are the Pauli matrices, $\mathbf{1}$ is an identity matrix and $\zeta$ represents an energy shift that can be neglected.  Thus $\langle  \mathbf{h}\rangle = \mathbb{A}\cdot\mathbf{S}_n$, where $\mathbf{S}$ is the (pseudo)spin of the particular doublet, and $\mathbb{A}$ is a $3\times 3$ matrix in real space.



The electronic wavefunctions are given by:
\begin{equation}
|\mathbf{k},n, s\rangle = \sum_{i,s} c_i^n (\mathbf{k}) |\mu_i,s\rangle,
\end{equation}
where $\mu_i$ indicates an atomic state with spin $s = \uparrow,\downarrow$, and $\mathbf{k}$ is the wavevector.  The  Ce 4f $j=5/2$  states are $|\mu_1,s\rangle = | \mp\frac{5}{2}\rangle$, $|\mu_2,s\rangle = |\mp\frac{1}{2}\rangle$, and $|\mu_3,s\rangle = |\pm\frac{3}{2}\rangle$; and the In 5p states are $|\mu_{4,5},s\rangle = | \pm 1, s\rangle$ (there is no $m_L = 0$ state considered). 
Note that the states $\mu_{1,2,3}$ are centered at the Ce site at $\mathbf{r} = (a/2, a/2)$, whereas  $\mu_{4,5}$  are centered at the In site at $\mathbf{r}=0$.  The $c_i^n(\mathbf{k})$ coefficients are determined by the eigenvectors of Eq. \ref{eqn:tightbinding}. In the $\mu_i$ basis $\mathcal{H}_e$ is given by \cite{Maehira2003}:
\begin{widetext}
 \begin{equation*}
\mathcal{H}_e =\left(
\begin{array}{ccccc}
  10 {B_2^0}+60 {B_4^0}+\frac{15 {V_{ff}}
   {\epsilon_{\mathbf{k}}}}{28} & -\frac{3}{14}
   \sqrt{\frac{5}{2}} {V_{ff}} {\gamma_{\mathbf{k}}}
   & 12 \sqrt{5} {B_4^4}+\frac{3}{28} \sqrt{5}
   {V_{ff}} {\epsilon_{\mathbf{k}}} &
   \sqrt{\frac{15}{7}} {c_{\mathbf{k}}} {V_{pf}} & -i
   \sqrt{\frac{15}{7}} {V_{\mathbf{k}}} {V_{pf}} \\
 -\frac{3}{14} \sqrt{\frac{5}{2}} {V_{ff}}
   {\gamma_{\mathbf{k}}} & -8 {B_2^0}+120
   {B_4^0}+\frac{3 {V_{ff}} {\epsilon
   k}}{14} & -\frac{3 {V_{ff}} {\gamma
   k}}{14 \sqrt{2}} & -i \sqrt{\frac{6}{7}}
   {V_{\mathbf{k}}} {V_{pf}} & -\sqrt{\frac{6}{7}}
   {c_{\mathbf{k}}} {V_{pf}} \\
 12 \sqrt{5} {B_4^4}+\frac{3}{28} \sqrt{5}
   {V_{ff}} {\epsilon_{\mathbf{k}}} & -\frac{3
   {V_{ff}} {\gamma_{\mathbf{k}}}}{14 \sqrt{2}} & -2
   {B_2^0}-180 {B_4^0}+\frac{3 {V_{ff}}
   {\epsilon_{\mathbf{k}}}}{28} & -\sqrt{\frac{3}{7}}
   {c_{\mathbf{k}}} {V_{pf}} & i \sqrt{\frac{3}{7}}
   {V_{\mathbf{k}}} {V_{pf}} \\
 \sqrt{\frac{15}{7}} {c_{\mathbf{k}}} {V_{pf}} & i
   \sqrt{\frac{6}{7}} {V_{\mathbf{k}}} {V_{pf}} &
   -\sqrt{\frac{3}{7}} {c_{\mathbf{k}}} {V_{pf}} &
   \Delta +{V_{pp}} {\epsilon_{\mathbf{k}}} &
   -{V_{pp}} {\gamma_{\mathbf{k}}} \\
 i \sqrt{\frac{15}{7}} {V_{\mathbf{k}}} {V_{pf}} &
   -\sqrt{\frac{6}{7}} {c_{\mathbf{k}}} {V_{pf}} & -i
   \sqrt{\frac{3}{7}} {V_{\mathbf{k}}} {V_{pf}} &
   -{V_{pp}} {\gamma_{\mathbf{k}}} & \Delta
   +{V_{pp}} {\epsilon_{\mathbf{k}}}
\end{array}
\right). 
\end{equation*}
\end{widetext}
We use $B_4^0 = -10.8$ K, $B_4^0 = 0.6$ K, $B_4^=1.5$ K as reported in Ref. \cite{WillersCEF115s} for CeRhIn$_5$,  and $V_{ff} = 4400$ K, $V_{pp} = 5730$ K, $V_{pf} = 5360$ K and $\Delta = 2300$ K  as reported in Ref. \cite{Maehira2003} for CeCoIn$_5$.   Studies were conducted by individually varying $B_2^0$, $B_4^0$, $V_{pf}$ and $\Delta$. In each case, the density of states, $N(E)$, was computed and the Fermi level, $E_F$, was determined such that the integrated density of states, $\int_{-\infty}^{E_F} N(E) dE$ remained constant. 

To determine the hyperfine field it is convenient to first compute the matrix elements of $h_{\alpha}$ in the $LS$ basis spanned by the states $|m_L^{4f},s\rangle$ with  $m_L^{4f} = -3, \cdots, +3$ for the Ce, and $|m_L^{5p},s\rangle$ with $m_L^{5p} = \pm 1$ for the In. In this case there are $7\times 2 + 2\times 2 = 18$ states, and the matrix elements can be written as:
\begin{widetext}
   \begin{equation}
  \langle m_L,s|h_\alpha|m_L',s'\rangle =  \langle m_L\left|\frac{L_\alpha}{r^3}\right|m_L'\rangle\delta_{ss'}+ \langle m_L\left|\sum F_{\alpha\beta}\right|m_L'\rangle\langle s| S_\beta|s'\rangle
\end{equation} 
\end{widetext}
where 
\begin{equation*}
  F_{\alpha\beta} = \frac{3x_\alpha x_\beta - r^2\delta_{\alpha\beta}}{r^5}.
\end{equation*}
The spatial wavefunctions for the Ce and In are given by:
\begin{eqnarray*}
  \langle \mathbf{r}-\mathbf{R}|m_L^{4f}\rangle &=& R_{4f}(r) Y_3^{m_L^{4f}}(\theta,\phi) \\
  \langle \mathbf{r}|m_L\rangle &=& R_{5p}(r) Y_1^{m_L^{5p}}(\theta,\phi)
\end{eqnarray*}
where $\mathbf{R} = (a/2,a/2, 0)$, and $a$ is the lattice constant. Here  $R_{nl}(r)$ are the hydrogenic  radial wavefunctions for the 4f and 5p states, with the effective nuclear charge $Z$ scaled such that the expectation values $\langle r \rangle$  equals the ionic radii of the Ce (1.15\AA) and the In (1.55\AA). Note that the $F_{\alpha\beta}$ elements involving the Ce sites must be computed in terms of $L_{\alpha} = -i\hbar (\mathbf{r}\times\nabla)_{\alpha}$ in cartesian coordinates because the Ce are displaced from the origin. The matrix elements $\langle m_L,s|h_\alpha|m_L',s'\rangle$ are computed numerically, and then the elements $\langle \mu_i,s|{h}_{\alpha}|\mu_j, s'\rangle$ are  obtained via a unitary transformation involving the appropriate Clebsch-Gordan coefficients and selecting the states corresponding to the $J=5/2$ manifold.  At this point there are a total of $3\times 2 + 2\times 2 = 10$ states.

The electronic Hamiltonian lifts the degeneracy of the $|\mu_i\rangle$ states and gives rise to a set of eigenstates $|\mathbf{k}, n\rangle$, where $n=1, ..., 5$ is a band index. Including spin degeneracy yields the states $|\mathbf{k}, n, s\rangle$.  The eigenvectors of $\mathcal{H}_e$ enable a unitary transformation of the matrix 
elements $\langle \mu_i,s|{h}_{\alpha}|\mu_j, s'\rangle$  to $\langle \mathbf{k}, n, s|{h}_{\alpha}|\mathbf{k}, n', s'\rangle$, which is a $10\times 10$ matrix.   We compute the $2\times 2$ matrix elements $\langle \mathbf{k}, n, s|{h}_{\alpha}|\mathbf{k}, n, s'\rangle$, for each state $|\mathbf{k},n\rangle$, and use Eq. \ref{eqn:hypmatrix} to determine $\mathbb{A}$. 

The matrix  $\langle \mathbf{k}, n|\mathbb{A}|\mathbf{k},n\rangle$ is then rotated and summed over the four nearest neighbor Ce sites at $\mathbf{R} = (a/2,a/2,0)$, $(a/2,-a/2,0)$, $(-a/2,a/2,0)$, and $(-a/2,-a/2,0)$, and summed over all states $|\mathbf{k},n\rangle$ at the Fermi level.  This matrix has dipolar symmetry, with the principal axis along the $z$ direction, hence is fully characterized by the quantity $A_{cc}$.

\bibliography{CeRhIn5pressure}
\end{document}